\documentclass[11pt]{article}
\usepackage{latexsym,amsmath,amsthm,amssymb,amsfonts}
\usepackage{amsbsy,wasysym,stmaryrd,hyperref,calrsfs,color,url,bbold,bm}
\usepackage{marginnote}
\usepackage[dvips]{graphicx,epsfig}
\csname @addtoreset\endcsname{equation}{section}
\definecolor{rougef}{rgb}{0.7,0,0}
\definecolor{vertf}{rgb}{0,0.6,0}
\definecolor{bleuf}{rgb}{0,0,0.95}

\newcommand{\be}{\begin{equation}}
\newcommand{\ee}{\end{equation}}
\newcommand{\bea}{\begin{eqnarray}}
\newcommand{\eea}{\end{eqnarray}}
\newcommand{\nn}{\nonumber}
\newcommand{\eq}[1]{(\ref{#1})}
\newcommand{\w}[1]{\\[0.#1cm]}
\newcommand{\cD}{{\cal D}}
\def\ds{\rule{0pt}{1.5ex}}

\begin{document}

\pagestyle{empty}
\begin{flushright}
  {  LMU-ASC 73/14}, MIFPA-14-40
\end{flushright}

\begin{center}
{\Large\bf  New Unfolded Higher Spin Systems in  $AdS_3$ }
\vspace{0.75cm}

{\bf Nicolas Boulanger${^{a}}$, Dmitry Ponomarev\,$^b$, Ergin Sezgin$^c$ and Per Sundell\,$^d$}\\

\vskip 15pt

{\em $^{a}$ \hskip -.1truecm
\textit{Service de M\'ecanique et Gravitation, Universit\'e de Mons --- UMONS\\ 20 Place du Parc, B-7000 Mons, Belgium}
\vskip 2pt }


\vskip 10pt

{\em $^b$ \hskip -.1truecm
\textit{Arnold Sommerfeld Center for Theoretical Physics,\\
Ludwig-Maximilians University Munich,\\
Theresienstr. 37, D-80333 Munich, Germany}
\vskip 2pt }


\vskip 12pt

{\em $^c$ \hskip -.1truecm George and Cynthia Woods Mitchell Institute for Fundamental Physics and Astronomy \\ Texas A\& M University, College Station,
TX 77843, USA
\vskip 2pt }


\vskip 12pt

{\em $^{d}$ \hskip -.1truecm
\textit{Departamento de Ciencias F\'isicas, Universidad Andres Bello,\\
Republica 220, Santiago, Chile}
\vskip 2pt }


\end{center}

\vspace{1cm}

\begin{minipage}{.90\textwidth}

\centerline{\textsc{Abstract} }

\vskip 5pt

We investigate the unfolded formulation of
bosonic Lorentz tensor fields of arbitrary spin
in $AdS_3$ containing a parity breaking mass parameter.
They include deformations of the linearisations of
the Prokushkin--Vasiliev higher spin theory
around its critical points.
They also provide unfolded formulations of
linearized topologically massive higher
spin fields including their critical versions.
The gauge invariant degrees of freedom are
captured by infinite towers of zero forms.
We also introduce two inequivalent sets of
gauge potentials given by trace constrained
Fronsdal fields and trace unconstrained
metric-like fields.

\end{minipage}

\vspace{.3cm}

\renewcommand{\thefootnote}{\arabic{footnote}}

\pagebreak

\pagestyle{plain}

\setcounter{footnote}{0}

\pagebreak



\section{Introduction}


Higher spin gravity theories in 3D have provided
a fruitful ground for exploring phenomena that have proven to be
difficult to tackle so far in higher dimensions, such as the nature
of black holes.
The existing 3D higher spin theories are essentially of two kinds.
The simpler and most studied one is purely topological, \emph{i.e.}
without any local bulk degrees of freedom, admitting a Chern--Simons
off-shell formulation \cite{Blencowe:1988gj}.
The second one, which was constructed by Prokushkin and Vasiliev
\cite{Prokushkin:1998bq}, has a finite number of propagating scalar
fields and is more akin to Vasiliev's 4D higher spin gravity.
What has been notably lacking, however, is an acceptable 
fully non-linear higher spin extension of topologically 
massive gravity, which provides a richer 3D toy model for 
quantum gravity as it propagates a massive graviton and 
has interesting black hole solutions of its own.
Proposals exist at the linearized \cite{Bagchi:2011td}
and the fully non-linear level \cite{Chen:2011yx}, but 
they are problematic as we shall comment on below,
and furthermore made within the standard 
approach to field theory that has not yet proven 
to be suitable for building up any fully nonlinear completion.

Indeed, all known fully non-linear formulations
of higher spin theories containing local bulk degrees of freedom
have been obtained within the unfolded approach to
field theory \cite{Vasiliev:1988xc,Vasiliev:1990vu,Vasiliev:1992gr}.
It has the characteristic feature of introducing one
infinite tower of auxiliary zero-form fields for
each local bulk degree of freedom.
These fields obey a chain of first-order equations
and as a result they encode all higher derivatives of the
dynamical field (in the metric-like formulation)  
that are non-vanishing on-shell.
In this chain, the dynamical equations reside 
at the lowest levels while the rest do not give 
any new information on-shell.
More precisely, in the case of a scalar field, the 
combination of the two lowest levels yields the
Klein--Gordon equation, while for spins $s\geqslant 1$
the dynamical equations reside at the first level,
in the form of Maxwell's equations for $s=1$
and their generalizations to spins $s\geqslant 2$
involving the generalized spin-$s$ Weyl tensors.

While seemingly uneconomic for lower-spin theories
with equations of motion that are second order in 
derivatives, unfolded dynamics becomes indispensable 
in the formulation of higher spin theories, as their
as Lorentz covariant interactions necessarily
involve higher derivatives.
Remarkably, as found by Vasiliev 
\cite{Vasiliev:1988xc,Vasiliev:1990vu,Vasiliev:1992gr},
unfolded dynamics not only provides a systematic
approach to higher spin interactions but also 
exhibits an underlying hidden symmetry in the
form of associative differential algebras.
This enabled Vasiliev to write down simple 
and elegant fully nonlinear equations in 4D 
in the form of a covariant constancy condition 
on a master zero-form and a constant-curvature 
condition on a master one-form.
Indeed, the 3D Prokushkin--Vasiliev (PV) theory \cite{Prokushkin:1998bq}
was formulated essentially along the aforementioned lines.
Its deformation to yield topologically massive 
fields with spins $s\geqslant 1$, however,
has turned out to be a challenging and still open problem.

Motivated by this, we here present an unfolded formulation of free 
topologically massive higher spin fields in $AdS_3$, that we expect 
will play a role in the fully nonlinear context, and that may also service as
a first step towards a classification of linearized such systems.
In particular, our construction gives a higher spin generalization
of various higher derivative extensions of gravity, including  
critical versions which harbor logarithmic modes; in the case of
gravity, see \cite{Bergshoeff:2010iy} for a review and references.

As noted above, a basic property of unfolded dynamics \cite{Vasiliev:1990vu,Vasiliev:1992gr,
Shaynkman:2000ts,Boulanger:2008up,Ponomarev:2010ji,
Didenko:2014dwa}
is that the local bulk degrees of freedom (but not boundary
states) are represented by infinite towers of gauge invariant 
Lorentz tensorial zero-forms that furnish infinite-dimensional 
representations of the isometry algebra of the spacetime background.
Thus, the unfolded analysis facilitates the characterization
of the propagating bulk modes using group theoretical methods
without any reference to gauge potentials.

In $AdS_3$, the propagating bulk modes can thus be identified via
the eigenvalues of the spin operators in the isometry algebra
${so}(2,2)\cong {so}(1,2)_{(+)} \oplus {so}(1,2)_{(-)}$, or equivalently,
in terms of the mass-square operator and the quadratic Casimir 
of ${so}(2,2)\,$\footnote{In \cite{Vasiliev:1992gr}, linearized
unfolded equations for tensor and spinor fields in $AdS_3$ were
constructed using spinor oscillators. In the present note, which
is limited to tensor fields, we shall use vector oscillators.}.
Although our main focus will be on unitary representations of lowest-energy
type, we shall also exhibit finite-dimensional representations and non-unitary
higher-order spin-s singletons as well as unitary representations in which the
energy is unbounded; for similar more generalized spectral analyses in $D\geqslant 4$, 
see \cite{Iazeolla:2008ix}.
We shall also extend the zero-form sector by elevating the
parity breaking mass parameter to an arbitrary matrix.
As we shall see, the extended system provides a unified description
of a number of higher spin theories, including theories that are 
topological, topologically massive and general massive and their  
critical versions (harboring logarithmic modes).

While the zero-form sector captures the gauge invariant local
bulk degrees of freedom, their interactions require gauge potentials, 
whose introduction at the linearized level is the topic of study
in Section \ref{sec:twisted}, where we make contact with
two different proposals given in the literature based on
either trace unconstrained metric-like potentials $h_{a_1\dots a_s}$ 
or Fronsdal potentials $\varphi_{a_1\dots a_s}$ .
Thus, the primary curvature tensor of the zero-form sector,
which is a divergence free traceless and symmetric 
Lorentz tensor $\Phi_{a_1\dots a_s}$, is equated to
either the dual of a de Wit--Freedman-type higher-spin 
curvature $R_{a_1 b_1|\dots | a_s b_s}$ (given by $s$ curls 
of $h_{a_1\dots a_s}$) or the Fronsdal curvature $F_{a_1\dots a_s}$
(built from up to two derivatives of $\varphi_{a_1\dots a_s}$) \cite{Manvelyan:2007hv,Bergshoeff:2009tb,Bergshoeff:2011pm}.
In other words, one and the same set of local degrees of freedom, 
carried by $\Phi_{a_1\dots a_s}$, can be integrated and 
carried by either $h_{a_1\dots a_s}$ or $\varphi_{a_1\dots a_s}$,
possibly leading to inequivalent non-linear extensions.

Turning to the existing proposals for topologically 
massive higher spins, a quadratic action has been given 
in \cite{Bagchi:2011td} and a fully nonlinear Chern--Simons 
action has been given in \cite{Chen:2011yx}.
Both yield linearized equations for spin $s\geqslant 2$ 
Fronsdal fields, which, however, lack trace conditions 
on the Fronsdal curvatures. 
The resulting traceless spin-$s$ modes and spin-$(s-2)$ 
trace modes come with opposite signs in the action, 
indicating the presence of ghosts.
In the unfolded formulation, however, the tracelessness of the 
curvatures is built into the equations of motion by construction.
Another difference is that the critical values for the mass parameter
$\mu$ at which the massive modes become singletons are spin independent 
in \cite{Bagchi:2011td,Chen:2011yx}, while they do depend on the spin
in our formalism.
Moreover, in both \cite{Bagchi:2011td} and \cite{Chen:2011yx} 
the aforementioned spin independence implies that the relative 
coefficient between the two terms in the equation of motion 
\eq{f1} below is given by $1/(\mu (s-1))$, which leads to decoupling
of spin-$1$ modes.
In the unfolded approach, this coefficient can be chosen freely,
and the natural choice is $1/(\mu s)$, allowing for spin-$1$
modes (but leading to spin dependent critical points).

As for the fully non-linear proposal in \cite{Chen:2011yx},
it has the additional problem (besides the presence of ghosts)
that the Cartan-Frobenius integrability conditions lead to 
nonlinear algebraic constraints on the fields on-shell. 
These constraints are satisfied at the linearized level
due to properties of the AdS$_3$ vacuum solution, but not 
at the second order and beyond, with the exception of special 
field configurations such as those considered in \cite{Chen:2012ana}.

The plan of the paper is as follows: In Section \ref{sec:twisted},
we construct the linear unfolded zero-form system in the case of a single
propagating degree of freedom by using Cartan-Frobenius
integrability to fix all free parameters in terms of the inverse
AdS radius $\lambda$ and a parity breaking mass-parameter, that we shall
denote by $\mu\,$.
We also exhibit an indecomposable structure, that arises for critical
values for $\mu$ as well as various dual representations parameterized
by two real numbers $\alpha,\beta\in [0,1]$,
and analyze the spectrum using harmonic expansion methods.
In Section~\ref{sec:Multiple}, we describe an extended zero-form
system consisting of $N$ unfolded towers mixed together via an
arbitrary mass matrix $\mu^i_j$ ($i,j=1,\dots,N$).
In Section \ref{sec:field}, we introduce the one-form gauge potentials in
the two aforementioned inequivalent fashions.
In the Conclusions, besides summarizing our results,
we also provide further comments on existing
proposals for quadratic \cite{Bagchi:2011td} and nonlinear
\cite{Chen:2011yx} actions for topologically massive higher
spin gravities, pointing out their problematic features.
Our conventions and some elements of ${so}(2,2)$ algebra
are given in Appendix \ref{sec:A}.
Appendix \ref{sec:B} contains technical details
related to the integrability of the unfolded equations
in the zero-form sector.


\section{Unfolded zero-form system: Single tower}
\label{sec:twisted}

In what follows we shall construct a general unfolded zero-form system
that describes a single spin-$s$ degree of freedom on an  AdS$_{3}$ background
(including the topological case which arises in the
limit where the mass parameter goes to infinity).
To this end, we write down an Ansatz for the unfolded equations for
an infinite set of totally-symmetric and traceless Lorentz tensors
\footnote{In 3D, the Lorentz algebra admits only such tensorial
representations, since any two anti-symmetrised indices can be
dualised to a (co)vector index using the epsilon tensor.
We use a notation wherein $\Phi_{a(n)}\equiv \Phi_{a_{1}\ldots a_{n}}$
and repeated indices denoted by the same letter
are implicitly symmetrized.
Symmetrization as well as anti-symmetrization is performed
with unit strength.}  $\{ \Phi_{a(n)}\}_{n\geqslant s}$.
All constants are fixed by the requirement of Cartan--Frobenius
integrability.
As a result, the couplings to the background frame field $h^a$
and the Lorentz connection in the Lorentz covariant derivative
$\nabla$ define a Lorentz covariant $so(2,2)$ representation matrix
for an infinite-dimensional $so(2,2)$ module, that we shall
denote by $\widetilde{\cal T}_{(s)}$ 
(see Appendix A for definitions and conventions). 
We shall then examine the spectrum of the system by expanding
the fields in terms of harmonic functions on $AdS_3$, which
is equivalent to expanding the Lorentz covariant basis elements
of $\widetilde{\cal T}_{(s)}$ in terms of compact basis elements
on which the two $so(2)$ subalgebras of $so(2,2)$ act diagonally.

\subsection{The unfolded equations}
%
%

Following Lopatin and Vasiliev \cite{Lopatin:1987hz}, we introduce vector oscillators obeying ($a=0,1,2\,$)
\begin{eqnarray}
[\alpha_a\ ,  \bar{\alpha}^b] = \delta^b_a\ ,
\label{oscill}
\end{eqnarray}
and collect the zero-forms in the master field\footnote{The operator $\bar{\alpha}$
has dimension of length and hence the operator $\alpha$ has dimension of mass, if no
dimensionful constant is introduced on the right-hand side of \eqref{oscill}.}
\begin{eqnarray}
\label{phi}
\vert \Phi^{(s)} \rangle   :=  \sum_{n=s}^{\infty} \tfrac{1}{n!}\,\Phi_{a(n)}\bar{\alpha}^{a_1}\ldots \bar{\alpha}^{a_n} \vert 0 \rangle \ .
\end{eqnarray}
Our Ansatz for the unfolded zero-form system reads ($n=s,s+1,...$)
\begin{eqnarray}
 \nabla \Phi_{a(n)} &=&
  h^a \,\Phi_{a(n+1)} +  \mu_n \,h^b\,\epsilon_{ba}{}^c \,\Phi_{a(n-1)c} +
\lambda_n \, \left[h_a \,\Phi_{a(n-1)} - \tfrac{n-1}{2n-1}\, h^b \,\Phi_{ba(n-2)} \,
\eta_{aa}\right]\ ,
\nonumber \\
&& \label{unfeq}
\end{eqnarray}
where $\mu_n $ and $\lambda_n$ are parameters of dimension mass and mass-squared,
respectively, and the last term on the right-hand side
removes the trace of $h_a \,\Phi_{a(n-1)}\,$ (so that the right-hand
side is symmetric and traceless).
In terms of the master-one form, the above equations assume the more compact form
\begin{eqnarray}
 {\cal D} |\Phi^{(s)} \rangle = 0\;,\quad
 {\cal D} := \nabla - {\rm i} \, h^a \tilde{\rho}(P_a) =
 \nabla - {\rm i} \,(\boldsymbol{\sigma}^-
            +\boldsymbol{\sigma}^0 +\boldsymbol{\sigma}^+) \ ,
\label{Unfolded}
\end{eqnarray}
where the operators  $\boldsymbol{\sigma}^-\,$, $\boldsymbol{\sigma}^0$ and
$\boldsymbol{\sigma}^+$ are given in Appendix \ref{sec:B}.
Eq. \eq{Unfolded} defines, via \eqref{unfeq}, the action of the
$AdS_{3}$ transvections $P_{a}$ in the Lorentz covariant basis
for the zero-form module $\widetilde{\cal T}_{(s)}$
(in which the Lorentz generators $M_{ab}$ act canonically).
As a result, the integrability condition ${\cal D}^{2}|\Phi^{(s)} \rangle=0$
endows $\widetilde{\cal T}_{(s)}$ with the structure of $so(2,2)$ module and
fixes the parameters $\mu_{n}$ and $\lambda_{n}\,$.

Before proceeding, we would like to remark that for $\mu_n=0$ and $s=0$, unfolded equations of 
the form \eq{unfeq} were considered in \cite{Vasiliev:1992gr} using (undeformed) spinor oscillators 
and shown to generically describe a massless scalar on $AdS_3$.
In \cite{Vasiliev:1992gr} also spinor fields on $AdS_3$ were considered,
for which $\mu_n$ is non-zero.
In \cite{Barabanshchikov:1996mc}, these systems were generalized by employing deformed
oscillator algebra so as to generically describe a massive scalars and fermions (except at critical points).
In \cite{Barabanshchikov:1996mc}, the significance of the restriction to $n\geqslant s$
at critical points was also discussed.
Subsequently, a nonlinear extension of this unfolded system, using
spinor oscillators, was obtained by Prokushkin and Vasiliev \cite{Prokushkin:1998bq}.
Unfolded equations of the
form \eq{unfeq} were studied by Lopatin and Vasiliev in
any dimension $D\geqslant 4$ \cite{Lopatin:1987hz}, using
vector oscillators, and in the absence of the $\mu_n$ term since these are tailored to $D=3\,$.

\paragraph{Analysis of Cartan-Frobenius integrability.} The details of the analysis of the integrability 
condition are given in Appendix \ref{sec:B}.
The required expressions for $\mu_{n}$ and $\lambda_{n}$ are
\begin{eqnarray}
{s=0}&:& \quad \mu_n = 0\;, \quad \forall\, n\in \mathbb{N}\ ,
\label{munpara1} \\
& &\nonumber \\
{s>0}&:& \quad \mu_n = \frac{\mu}{n+1}\ , \quad  n\in \mathbb{N}\ ,
\quad n\geqslant s \ ,
\label{munpara2}
\end{eqnarray}
and
\begin{eqnarray}
{s=0}&:& \quad \lambda_n = \frac{n \lambda^2}{2n+1}\left[
\frac{M_0^2}{\lambda^2} +1 -  n^2\right]\ ,\qquad n \in \mathbb{N}\ ,
\label{lambdan1}\\
& & \nonumber \\
\label{sge0}\label{lambdan2}
{s>0}&:& \quad \lambda_n = \frac{n^2-s^2}{n(2n+1)}
\left[\frac{\mu^2}{s^2} -  \lambda^2\,n^2\right]\ , \quad  n\in \mathbb{N}\ , \quad
n \geqslant s \ ,
\end{eqnarray}
where where $\mu$ and $M_0$ are arbitrary real parameters with dimension of mass.
Thus the unfolded equations in $\widetilde{\cal T}_{(0)}$ take the form ($n \geqslant 0$)

\be
\nabla_b \Phi_{a_1...a_n} = \Phi_{b a_1...a_n}
 + \frac{n\,\lambda^2}{2n+1} \left(\frac{M_0^2}{\lambda^2} +1 -  n^2\right)\,
     \left( \eta_{b(a_{1}} \,\Phi_{a_{2}\ldots a_n)}- \frac{n-1}{2n-1}\;
 \eta_{(a_1a_2} \Phi_{a_3...a_n)b}\right)  \ ,
 \label{m1}
\ee
while for $s>0$ they take the following form in $\widetilde{\cal T}_{(s)}$ ($n \geqslant s$):
\begin{eqnarray}
\nabla_b \Phi_{a_1...a_n} &=&   \Phi_{b a_1...a_n}  +\frac{\mu}{n+1}\,\epsilon_{b(a_1}{}^c\, \Phi_{a_2\cdots a_n)c}
\label{m2}\w2
 && +\;  \frac{n^2-s^2}{n(2n+1)} \left[\frac{\mu^2}{s^2} -  \lambda^2\,n^2\right]
  \left( \eta_{b(a_{1}} \,\Phi_{a_{2}\ldots a_n)}  - \frac{n-1}{2n-1}\;
 \eta_{(a_1a_2} \Phi_{a_3...a_n)b}\right)  \ .
 \nonumber
\end{eqnarray}
For $\mu=0$, the latter equation system coincides with the linearized
equations in the zero-form sector of the PV theory
at the critical point where the deformation parameter $\nu=2s+1$ \cite{Barabanshchikov:1996mc,Prokushkin:1998bq} (see \eq{v3} below).

\paragraph{Critical points.} From (\ref{lambdan2}) it follows that
if $\mu=\pm \, \lambda \,s\,s'\,$ for some integer $s' > s\,$,
then $\lambda_{s'}$ vanishes and hence the set
$\{ \Phi_{a(n)} \}_{n\geqslant s'}$ remains invariant
under covariant differentiation\footnote{
Such a phenomenon also appears for massive scalars \cite{Shaynkman:2000ts} and
higher spin fields \cite{Ponomarev:2010st} for special values of the mass.
More precisely, in the limit when the mass of a spin-$s$
field in $AdS_d$ tends to a specific value $m_t$ it decomposes
into a direct sum of a partially massless field of spin $s$ and
depth $t$ and a massive field of spin $t$. The zero-form module
of the latter field contains a finite-dimensional submodule.
Such modules are also employed in unfolding as  modules where
gauge potentials take their values.}.
Thus, at these critical points, the $so(2,2)$ module
$\widetilde{\cal T}_{(s)}$ exhibits an indecomposable
structure.
As we shall see in Section 2.2, the finite-dimensional representations
$\{\Phi_{a(n)}\}_{n=s,\dots,s'-1}$ show up in the spectral 
analysis.

\paragraph{The Casimir operators.}
Using the unfolded equations, we can evaluate the Casimir
operator $C_{2}[so(2,2)]$ in $\widetilde{\cal T}_{(s)}\,$.
From \eqref{Unfolded}, (\ref{munpara2}) and (\ref{lambdan2}) it follows
that the value of the mass-squared operator $M^{2} := - \eta_{ab}P^aP^b$
on the primary zero-form $\Phi_{a(s)}$ in $\widetilde{\cal T}_{(s)}$ is
given by
\begin{eqnarray}
\tilde{\rho}(- P^bP_b)\Phi_{a(s)}=\left(\frac{\mu^2}{s^2}-\lambda^2\,(s+1)\right)\Phi_{a(s)}\ ,
\quad s>0\ .
\label{p2}
\end{eqnarray}
Thus $\Phi_{a(s)}$ obeys\footnote{Note that this equation can also be obtained by acting with  
$\epsilon_{\nu_1}{}^{\mu\nu}\,\nabla_\mu$ on both side of the fist order equation \eq{tmg}.}
\begin{eqnarray}
\left[ \Box - \left(\frac{\mu^2}{s^2}-\lambda^2\,(s+1) \right) \right] \Phi_{a(s)} = 0\ .
\label{qe}
\end{eqnarray}
Using also the fact that
\begin{eqnarray}
\tilde{\rho}\left(\frac{1}{2}M^{ab}M_{ab}\right)\Phi_{a(s)}=s(s+1)\,\Phi_{a(s)}\ ,
\end{eqnarray}
it follows that for $s>0$
\begin{eqnarray}
C_2[so(2,2)]\vert_{\widetilde{\cal T}_{(s)}}=s^2-1+\frac{\mu^2}{\lambda^2\,s^2}\ .
\label{Casimirso22}
\end{eqnarray}
In the case of $s=0$, we find
\be 
C_2[so(2,2)]|_{\widetilde{\cal T}_{(0)}}= \lambda^{-2} M_0^2\ .
\ee
The algebra ${so}(2,2)={so}(1,2)_{(+)}\oplus so(1,2)_{(-)}$ generated by
\bea
& J^{(\varepsilon)}_a := \tfrac{1}{2}\, (M_a + \frac{\varepsilon}{\lambda} P_a\,)\ ,\quad
M^{a}~:=~\tfrac{1}{2}\, \epsilon^{abc} M_{bc}\ ,\quad \epsilon^{012} = 1\ ,&
\nonumber\\
& [ J^{(\varepsilon)}_a,J^{(\varepsilon)}_b ] = (-{\rm i})\,\epsilon_{abc} \,\eta^{cd} J^{(\varepsilon)}_d\ ,\quad
[J^{(+)}_a,J^{(-)}_b]~=~0\ .&
\eea
The corresponding Casimir operators
\be
C_2[{so}(1,2)_{(\varepsilon))} ] =
-\tfrac{1}{4}\, \Big( C_2[{so}(1,2)_{\rm Lor}] - \frac{P^a P_a}{\lambda^{2}}\,\Big) 
+ \frac{{\varepsilon}}{2\lambda} M^a P_a\  ,\ee
assume the following values in $\widetilde{\cal T}_{(s)}$ for $s>0$:
\be C_2[{so}(1,2)_{(\varepsilon))} ]|_{\widetilde{\cal T}_{(s)}}=- \frac{1}{4}\, 
\left( s^2 - 1 + \frac{\mu^2}{\lambda^{2}s^2}\, \right) + \frac{\varepsilon\mu}{2\lambda}\ ,
 \label{c2pm}
\ee
where we have used \eq{unfeq}, \eq{p2} and \eq{vectorrepLor}.

\paragraph{Primary equations of motion.}
%
Combining the equations \eq{m1} for $n=0$ and $n=1$ yields the scalar field equation
\be
\left(\Box - M_0^2  \right)\Phi  = 0\ .
\label{scalar}
\ee
The first equation in the chain \eq{m2}, on the other hand, reads
\bea
\nabla_{\mu} \Phi_{\nu(s)}=\Phi_{\mu \nu(s)}+\frac{\mu}{s+1}\;\epsilon_{\mu \nu}{}^\rho\,
\Phi_{\nu(s-1)\rho}\ .
\label{pp}
\eea
Taking a trace yields
\be
\nabla^\mu \Phi_{\mu \nu(s-1)} = 0 \  .
\label{div1}
\ee
Contracting with $\epsilon^{\lambda\mu\nu}$  produces
\be
\Phi_{\nu(s)} + \frac{s}{\mu} \;\epsilon_{\nu}{}^{\rho\sigma}\,\nabla_{\rho}\Phi_{\sigma\nu(s-1)}= 0\ ,\qquad s>0\ ,
\label{tmg}
\ee
where the second term is totally symmetric by virtue of (\ref{div1}).
For $s=2$ and and appropriate identification of $\mu$, Eq. \eqref{tmg} 
is the linearized field equation of topologically massive gravity 
\cite{Deser:1981wh}.
Upon introducing gauge potentials (see in Section \ref{sec:field}), 
the tensor $\Phi_{\nu(s)}$ is identified with either
the dual of the de Wit--Freedman-like curvature or the 
Fronsdal tensor.
In the first case, \eq{tmg} becomes an $(s+1)$-derivative field equation for
a metric-like gauge potentials, while in the latter
case it takes the form of a three-derivative field equation for a 
Fronsdal gauge potential.

\paragraph{Limits.} The field equation \eq{tmg} has  the following two noteworthy limits:
\bea
\mu \to \infty\ :  \quad && \Phi_{\nu(s)}=0\ , \qquad s>0\ ,
\label{L1}\w2
\mu \to 0\ : \quad && \epsilon_{\nu}{}^{\rho\sigma}\,\nabla_{\rho}\Phi_{\sigma\nu(s-1)}= 0\ ,\qquad s>0\ .
\label{L2}
\eea
In the $\mu\to \infty$ limit, all the spin $s>0$ fields become  
nonpropagating.
Upon the introduction of Fronsdal gauge potentials,
this case corresponds to the linearized 
Blencowe theory \cite{Blencowe:1988gj}, 
or, equivalently, the spin $s>0$ sector of the linearized
PV system for generic values of the
deformation parameter $\nu$ \cite{Vasiliev:1992ix,Prokushkin:1998bq}.

The $\mu\to 0$ limit corresponds to the
zero-form field equations of the linearized critical
PV system at $\nu=2s+1$.
At these critical points, the linearization of the
PV curvature constraint, however,
becomes a more subtle matter due to the fact that the
source term in this equation becomes singular in the critical limits.
It remains to be seen if a suitable regularization of
this term can lead to the equations that we shall propose
in Section \ref{sec:field}.
If so, then Eq. \eq{L2}, with $\Phi_{a(s)}$ now expressed
in terms of potentials, becomes the linearized field equation
for conformal Chern--Simons gravity in 3D for the case of spin 2 \cite{Deser:1981wh,Horne:1988jf}, which has been
studied in  detail  in \cite{Afshar:2011qw}.
For $s>2$, however, it describes either an $s+1$ or three derivative
field equation depending on the nature of the gauge potential introduced.
We also note that such a tentative equation would differ from the
vanishing of the spin-$s$ Cotton tensor, which is expected to involve $2s-1$
derivatives \cite{Pope:1989vj,Bergshoeff:2011pm,Nilsson:2013tva}.
Nonetheless the prospects of their exhibiting similar behaviour, owing to
higher-spin symmetries (as opposed to Weyl symmetry) has been noted in \cite{Afshar:2011qw}.

\paragraph{Indecomposable structures.}

The unfolded system \eqref{Unfolded} remains integrable after one performs
the following rescaling:
\begin{eqnarray}
&\boldsymbol{\sigma}^-_{\rm new} = {\cal N}(N+1)\,\boldsymbol{\sigma}^-_{\rm old} ~,
\qquad
\boldsymbol{\sigma}^+_{\rm new} =  \Theta(N-s)
{{\cal N}^{-1}(N)}\,\boldsymbol{\sigma}^+_{\rm old} ~ ,&
\nonumber \\
& \Theta(x) = 0 \quad {\rm for}\quad x\leqslant 0\quad {\rm and}\quad
 \Theta(x) = 1 \quad {\rm for}\quad x > 0\ , &
\label{B11}
\end{eqnarray}
provided ${\cal N}(n)$ and ${\cal N}^{-1}(n) \Lambda(n)$ are non-vanishing.
In other words, the system \eqref{unfeq} remains integrable upon
multiplication of the first term on the right-hand side by
${\cal N}_{n+1}$ and sending $\lambda_n$ to $\lambda_n/{\cal N}_n$.
These rescalings can be used to alter the indecomposable structure of
$\widetilde{\cal T}_{(s)}$ at the critical points $\mu = \lambda s s'\,$.
Letting $\alpha,\beta\in [0,1]$, the general (in)decomposable
structures for $s>0$ can be obtained from
\bea
\nabla_b \Phi_{a_1...a_n} &=&  \left[\frac{(n+1)^2-s^2}{(2n+1)(2n+3)}\right]^{\alpha}\,
\left[\frac{\mu^2}{s^2} -  \lambda^2\,[n+1]^2\right]^{\beta}\Phi_{b a_1...a_n}
+\frac{\mu}{n+1}\,\epsilon_{b(a_1}{}^c\, \Phi_{a_2\cdots a_n)c}
\nn \w2
 && +\;  \Theta(n-s)\,\frac{2n-1}{n}\, \left[\frac{n^2-s^2}{(2n-1)(2n+1)}\right]^{1-\alpha}\,
 \left[\frac{\mu^2}{s^2} -  \lambda^2\,n^2\right]^{1-\beta}\,  \times
 \nn \\
 &&\qquad \quad
 \times \left( \eta_{b(a_{1}} \,\Phi_{a_{2}\ldots a_n)}  - \frac{n-1}{2n-1}\;
 \eta_{(a_1a_2} \Phi_{a_3...a_n)b}\right)  \ ,
 \label{m2bis}
\eea
where we have chosen a nonsingular $n$ dependent factor in ${\cal N}_n$
for the convenience of comparing our result with that of \cite{Barabanshchikov:1996mc,Prokushkin:1998bq}.
Indeed, for $\mu=0$ and using the  terminology employed in
\cite{Barabanshchikov:1996mc} (see \eq{v4} below),
we see that $\alpha=\beta=0$ gives the co-twisted representation,
while $\alpha=1,\beta=0$ gives the (dual) twisted representation
\cite{Barabanshchikov:1996mc,Prokushkin:1998bq}.
For $\mu\ne 0$, the choice $\alpha=0, \beta=0$ gives our original
unfolded system \eq{unfeq} whose indecomposable structure at the
critical values for $\mu$ was spelled out above, while other choices yields
other inequivalent indecomposable structures at these critical points.
Thus, for $\beta=0$ or $\beta=1$, the ideals closed under
exterior covariant differentiation are given by
$\{\Phi_{a(n)}\}$ for $n\ge s'$ and $n \le s'-1$, respectively.

\paragraph{Multi-spinor notation.} We conclude this section by
converting the unfolded equations from vector indices to two-component
Majorana spinor indices,
as to make more explicit contact with the linearized critical PV 
system \cite{Barabanshchikov:1996mc,Prokushkin:1998bq}.
To this end, we employ ${so}(1,2)$ Dirac $\gamma$-matrices to define
\be
 V_a =  \frac{\rm i}{\sqrt 2}\,  \left(\gamma_a\right)^{\alpha\beta}\,V_{\alpha\beta}\ ,\qquad
 V_{\alpha\beta} = \frac{\rm i}{\sqrt 2}\, V^a \left(\gamma_a\right)_{\alpha\beta}\ , \qquad
\gamma^a_{(\alpha\beta}\,\gamma^b_{\gamma)\delta}\,\eta_{ab} =0\ .
\ee
Thus, the $\mu$-deformed unfolded system \eq{m2}  in spinor notation takes the form
\bea
\nabla_a \Phi_{\alpha_1...\alpha_{2n} }&=&  \frac{\rm i}{\sqrt 2} 
\Bigg[ \left(\gamma_a\right)^{\beta\gamma}\,\Phi_{\beta\gamma \alpha_1 ... \alpha_{2n} }
-\frac{{\rm i}\mu}{4{\sqrt 2} (n+1)}\, \left(\gamma_a\right)_{(\alpha_1}{}^\beta\,
\Phi_{\alpha_2...\alpha_{2n})\beta}
\nn\w2
&& \qquad +  \frac{n^2-s^2}{n(2n+1)} \left[\frac{\mu^2}{s^2} -  \lambda^2\,n^2\right]
\left( \gamma_a \right)_{(\alpha_1\alpha_2}\, \Phi_{\alpha_3...\alpha_{2n)} } \Bigg]\ ,
\label{v3}
\eea
while the dual unfolded system \eqref{m2bis} with $\alpha=1$ and $\beta=0$ reads
\bea
\nabla_a {\widetilde\Phi}_{\alpha_1...\alpha_{2n} }&=&
\frac{\rm i}{\sqrt 2} \Bigg[  \frac{(n+1)^2-s^2}{(2n+1)(2n+3)}  {\widetilde \Phi}_{a \alpha_1 ... \alpha_{2n} }
-\frac{{\rm i}\mu}{4{\sqrt 2} (n+1)}\, \, \left(\gamma_a\right)_{(\alpha_1}{}^\beta\,
{\widetilde\Phi}_{\alpha_2...\alpha_{2n})\beta}
\nn\w2
&&\qquad  +   \frac{2n-1}{n} \left[\frac{\mu^2}{s^2} -  \lambda^2\,n^2\right]
 \left( \gamma_a \right)_{(\alpha_1\alpha_2}\, {\widetilde\Phi}_{\alpha_3...\alpha_{2n)} }\Bigg]\ .
 \label{v4}
\eea
%

 \subsection{ Spectrum and unitarity}

The question of spectrum depends  on which representations of ${so}(2,2)$ we employ in
analyzing the field equations \eq{scalar} and \eq{tmg}. It is an easy matter to determine
the spectrum if we employ the lowest-weight
UIRs that have the property of being finite at the origin and having finite norm with
respect to the group invariant Haar measure (see, for example, \cite{Deger:1998nm,rnn}).
They are denoted by   $D(E_0,s_0)$ where $E_0$ is the lowest energy and
$s_0$ is the helicity of the lowest energy state, which must be an integer.
In this subsection we shall take $\mu\ge 0$, without any loss generality, and  we shall  set
the inverse $AdS_3$ radius $\lambda=1$.

\paragraph{Unitary representations.}

In the case of the scalar field, its lowest-energy representations are $D(1\pm\sqrt{1+M_0^2}, 0)\,$, where
reality of $E_{0}$ requires  the Breitenlohner--Freedman bound $M_{0}^{2}\geqslant -1$. %
The irrep with the upper sign is unitary, while the lower sign
case is unitary provided that  $-1\leqslant M^{2}_{0} \leqslant 0\,$.

Turning to \eq{tmg}, we expand the primary zero-form as
\be
\Phi_{a(s)} (x) =\sum_{q} \Phi_q^{(E_0,s_0)}\, Y_{a(s),q}^{(E_0,s_0)} (x) \ ,\quad E_0 \geqslant |s_0|\ ,
\label{he}
\ee
where $\Phi_q^{(E_0,s_0)}$ are constants and
$Y_{a(s),q}^{(E_0,s_0)}(x)$ are totally symmetric
and divergence-free ${so}(2,2)$ representation functions
in $D(E_0,s_0)$ obeying
\footnote{Lorentz invariance dictates this form of the result, and the coefficient is easily obtained by acting with
$\epsilon^{\mu\nu a}\,\nabla_\nu$  on both sides of the equation, then using \eqref{A14}.}
\be
\epsilon_{\ds a}{}^{bc} \nabla_{\ds b} Y^{(E_0,s_0)} _{a(s-1)c,q}
=  -{\rm sign}(s_0) (E_0-1)\, Y^{(E_0,s_0)}_{a(s),q}\ ,
\label{hexp}
\ee
and the sum includes all lowest-weight representations that
contains a symmetric and traceless Lorentz spin-$s$ tensor, which 
implies $s_0=\pm s$.
Substituting this expansion into \eq{tmg}, one finds for $\mu>0$ that
\footnote{Choosing $\mu<0$ gives the opposite helicity
solution with lowest weight $(1-\frac{\mu}{s},-s )$. }
\be
(E_{0},s_0) = \left(1 + \frac{\mu}{ s} ,\, s\right) \ ,   \qquad s\geqslant 1\ ,\qquad \mu \ge s(s-1)\ ,
\label{irrep}
\ee
where the unitarity bound is saturated by singleton representations.
For $\mu=0$, there are two possibilities, namely the singleton representations
$D(1,\pm 1)$.


\paragraph{Compact basis.}

In order to exhibit the ${so}(1,2)_{(+)} \oplus {so}(1,2)_{(+)} $ content of the spectrum, we use \eq{a12}
and introduce compact basis vectors $|m^{(+)},m^{(-)}\rangle$  for $\widetilde {\cal T}_{(s)}$
obeying\footnote{We assume that the multiplicity for each $(m^{(+)},m^{(-)})$ vector to be $0$ or $1$.}
\begin{equation}
(J^{(\varepsilon)}_0 - m^{(\varepsilon)})\;|m^{(+)},m^{(-)}\rangle ~=~0\ .
\end{equation}
It is important to note that $m^{(+)} + m^{(-)}$ must be integer in order for Lorentz tensors to be expandable in this basis.
As can be seen from
\begin{eqnarray}
2 {\rm i}\,J^{(\pm)}_{-} = -i L^{\mp}_{2} + L^{\mp}_{1}\;,\qquad
2 {\rm i}\,J^{(\pm)}_{+} = -i L^{\pm}_{2} - L^{\pm}_{1}\ ,
\end{eqnarray}
the lowest energy representations are given by direct products of  highest weight
representations of ${so}(1,2)_{(-)}$ with  lowest weight representations of ${so}(1,2)_{(+)}$.
Using
\be
C_2[{so}(1,2)_{(\varepsilon)} ] = J^{(\varepsilon)}_{\varepsilon}  J^{(\varepsilon)}_{-\varepsilon}
- J^{(\varepsilon)} _0 ( J^{(\varepsilon)}_0 - \varepsilon )\ ,
\ee
and \eq{c2pm}, one finds the characteristic equation ($s>0\,$)
\begin{equation}
-j^{(\varepsilon)} \,(\,j^{(\varepsilon)} - \varepsilon \, ) ~=~ - \frac{1}{4}\,
\left( s^2 - 1 + \frac{\mu^2}{s^2}\, \right) + \frac12 {\varepsilon\mu}\ ,
\end{equation}
where $j^{(\varepsilon)}$ are the eigenvalues of $J_0^{(\varepsilon)}$, with roots
\begin{equation}
j_{\pm}^{(\varepsilon)} ~=~ \tfrac{1}{2}\, \left(\, \varepsilon \pm \left[ s -  \,
\frac{\varepsilon\mu}{ s}\right]\,\right)\ .
\label{solutionforj}
\end{equation}
The condition of integer spin means that the allowed roots are  $(j^{(+)}_{-},j^{(-)}_{-})$  corresponding to
$D(1+\frac{\mu}{s} , s)$ which is unitary for
$\mu\ge s(s-1)$,  and $(j^{(+)}_{+},j^{(-)}_{+})$ corresponding to  $D(1 - \frac{\mu}{s} , - s)\,$ which is
unitary for $\mu \le s(1-s)$. These  unitarity conditions are  clearly seen from those on the representations
of the ${so}(1,2)$ subalgebras
\footnote{ The lowest or highest-weight unitary and irreducible representations of $so(1,2)$ are
the infinite-dimensional discrete series $D^+( j^{+})$ and $D^-(j ^{-})$
that are respectively lowest-weight and highest-weight and where $j^{+}  > 0\;$, $j^{-}  < 0\;$,
plus  the trivial, 1-dimensional representation $j=0\,$. }.
%

\paragraph{Finite dimensional representations.}
%
At the critical points
\begin{equation}
\mu = \pm  s\, s' \ ,\qquad   s'=s+1,s+2,...  \,
\label{dp}
\end{equation}
finite dimensional representations arise. These are given by  direct products of finite dimensional ${so}(1,2)_{(+)}$
representations with negative lowest weights
$j^{(+)}_{-} = \tfrac{1}{2}\,(1+ s - s')$ with finite dimensional $so(1,2)_{(-)}$
representations with positive weights $j^{(-)}_{+} = \tfrac{1}{2}\,(-1+s +s')\,$. The finite dimensions are given by
 $s'^{2} - s^{2}$, which are indeed equal to the sum of the dimensions of the Lorentz tensors
$\{ \Phi_{a(n)}\}_{n = s, s+1, \ldots, s'-1}\,$.

\paragraph{Higher-order singletons.}
%
At the degenerate points  \eq{dp}, there also arise interesting infinite dimensional lowest energy representations. They are given by
direct products of infinite-dimensional $so(1,2)_{(-)}$ representations with negative highest weights $j^{(-)}_{-} = - (s +s'+1)/2$
with  finite-dimensional $so(1,2)_{(+)}$ representations with negative lowest weights $j^{(+)}_{-} = (-s'+s+1)/2\,$.
Correspondingly, we have
\begin{equation}
(E_0,s_0) = (s+1, s')\ ,
\end{equation}
which is unitary only for $s'=s+1$, yielding the singleton $D(s+1,s+1)$
\footnote{See \cite{Basile:2014wua}  for a recent work where tensor products of higher-order singletons
were considered via  character formulae. Some earlier works involving higher order singletons include
\cite{Iazeolla:2008ix,Boulanger:2008kw} and \cite{Bekaert:2013zya}.}.

\paragraph{Representations with unbounded energy.} Finally, let us mention that we could relax the condition of lowest-energy for
a unitarizable $so(2,2)$ module by building 3D analogs of the higher-dimensional UIRs
found in \cite{Iazeolla:2008ix}. These involve the principal and complementary
series of $so(1,2)_{(\pm)}\,$, and they are built by taking direct products of the representations
of ${so}(1,2)_{(\varepsilon)}$ both or which are lowest-weight or highest-weight type.
At the classical level, and in $D\ge 4$, it was argued \cite{Iazeolla:2008ix} that they give fully nonlinear
solutions that are higher spin analogs of the soliton in 5D \cite{Horowitz:1998ha},
but it remains an open question whether they actually give rise to unitary states in the
quantum theory.

\section{Unfolded zero-form system: Multiple towers}
\label{sec:Multiple}

Let us extend the previous construction to a more
general setting. From now on we will use a set of zero
forms $\{\Phi^{i}_{a(n)}\}_{n=s, s+1,\ldots}$ enumerated
by an index $i=1, \ldots , N\,$.
All previous considerations remain valid with the sole difference that
$\mu_n  $ and $\lambda_n $ become matrix valued functions
of $n$: $\mu_n \rightarrow (\mu^i{}_j)_n$ and $\lambda_n \rightarrow (\lambda^i{}_j)_n$.
Integrability of the unfolded equations requires the $\mu$ matrix to commute with
$\lambda$ matrix, and  the general solution of the integrability condition is given by
\begin{eqnarray}
{s=0}&:& \quad (\mu^i{}_j)_n = 0 \quad \forall n\in \mathbb{N}\ ,
\\
& & \nonumber \\
{s>0}&:& \quad (\mu^i{}_j)_n = \frac{\mu^i{}_j}{n+1} \quad  n\in \mathbb{N}\ ,
\quad n\geqslant s \ ,
\\
& & \nonumber \\
{s=0}&:& \quad (\lambda^i{}_j)_n = \frac{n\,\lambda^{2}}{2n+1} \left[
\frac{(M_0^2)^i{}_j}{\lambda^2} +(1 -  n^2)\;\delta^i{}_j\right]\;,\qquad
n \in \mathbb{N}\ ,
\\
& & \nonumber \\
{s>0}&:& \quad (\lambda^i{}_j)_n = \frac{n^2-s^2}{n(2n+1)}
\left[\frac{(\mu^2)^i{}_j}{ s^2} -  \lambda^{2\,}n^2\;\delta^i{}_j\right]\ ,  \quad  n\in \mathbb{N}\ , \quad
n \ge s \ . \\
& & \nonumber
\end{eqnarray}
It follows that
\begin{eqnarray}
\rho(- P^bP_b)\,\Phi^j_{a(s)}=\left(\frac{(\mu^2)^j{}_i}{s^2}-\lambda^2 (s+1)\;
\delta^j{}_i\right)\Phi^i_{a(s)}\ ,
\end{eqnarray}
which means that the mass-squared operator is not necessarily diagonal. Only the matrix
 $\mu^i{}_j$ arises in the linearized field equations of the primary fields,
 which  take the form
\bea
&&
\left(\Box \delta^i_j - (M_0^2)^i{}_j  \right)\Phi^j  = 0\ ,
\label{ne1}\w2
&&
\nabla_{\mu} \Phi^i_{\nu(s)}=\Phi^i_{\mu \nu(s)}+\frac{\mu^i{}_j}{s+1}\;\epsilon_{\mu \nu}{}^\rho\,
\Phi^j_{\nu(s-1)\rho}\ .
\label{pp2}
\eea
As before, it follows that
\be
\frac{1}{s}\,\mu^i{}_j\, \Phi^j_{\nu(s)} + \epsilon_{\nu}{}^{\rho\sigma}\,\nabla_{\rho}\Phi^i_{\sigma\nu(s-1)}= 0\ ,\qquad
\nabla^\mu \Phi^i_{\mu \nu(s-1)} = 0 \  ,\quad s>0\ .
\label{ne2}
\ee
Generally the matrix $\mu^i{}_j$ can always be brought to Jordan form, which we shall consider shortly. As a simple example,
however, let us take the matrix $\mu^i{}_j$ to be
\be
\mu^i{}_j  = \begin{pmatrix} 0& m\\ m &0\end{pmatrix}
\ee
where $m$ is an arbitrary constant with dimension of mass.
Substitution into \eq{ne2} gives
\begin{eqnarray}
\label{nmg1}
\epsilon_{\nu}{}^{\rho\sigma}\, \nabla_{\rho}\Phi^{(1)}_{\sigma\nu(s-1)}
 + \frac{m}{s} \; \Phi^{(2)}_{\nu(s)} &=& 0\ ,
\w2
\label{nmg2}
\epsilon_{\nu}{}^{\rho\sigma}\,\nabla_{\rho}\Phi^{(2)}_{\sigma\nu(s-1)}
+ \frac{m}{s}\; \Phi^{(1)}_{\nu(s)} &=& 0 \ .
\end{eqnarray}
Expressing $\Phi^{(2)}$  in terms of $\Phi^{(1)} \equiv \Phi$ from (\ref{nmg1}) and plugging
the result into (\ref{nmg2}) yields
\bea
\Box \Phi_{\nu(s)} - \left( \frac{m^2}{s^2} - \lambda^2 (s+1)\right)
\Phi_{\nu(s)} &=& 0\ .
\label{nmg}
\eea
For $s=2$, this equation, with appropriate identification of the mass parameters, is the linearized 
field equation of new massive gravity \cite{Bergshoeff:2009hq}. The equation can be factorized as
\be
\left[  \cD \left(\tfrac{\lambda s}{m}\right) \cD \left(-\tfrac{\lambda s}{m}\right) \Phi^{(s)} \right]_{\nu(s)}=0\ ,
\label{fe}
\ee
where ${\cal D}(\eta)$ are first-order linear differential operators defined by \cite{Bergshoeff:2010iy}
\be
 \left[ {\cal
D}(\eta) \Phi^{(s)}\right]_{\nu(s)} = \left[{\cal
D}(\eta)\right]_{\nu_1}{}^\rho \Phi_{\rho\nu_2 \cdots \nu_s}\ ,
\qquad
\left[ \mathcal{D}\left(\eta\right)\right]_\mu{}^\nu := \lambda
\delta_\mu^\nu + \frac{\eta}{\sqrt{|\bar g|}}\varepsilon_\mu{}^{\tau\nu}\nabla_\tau\ .
\label{def}
\ee
The rank-$s$ tensor $\cD(\eta)\Phi^{(s)}$  is traceless, totally symmetric  and divergence-free
provided $\Phi$ satisfies the same constraints. From \eq{fe} we see that the eigenfunctions of $\Box$ acting 
on spin-$s$ fields are linear combinations of
solutions to the equations
\be
\cD \left(\tfrac{\lambda s}{m}\right) \Phi^{(s)}_{\pm} =0 \ ,
\qquad
\cD \left(\tfrac{-\lambda s}{m}\right) \Phi^{(s)}_{\pm} =0\ .
\ee
These are, of course,  the equations \eq{nmg1} and \eq{nmg2} for the combinations $\Phi^{(s)}_\pm := \Phi^{(s)}_1 +\Phi^{(s)}_2$.

Let us now consider the case in which $(\mu)^i{}_j$ has $N$ distinct non-vanishing
eigenvalues, \emph{viz.}
\be
(\mu)^i{}_j = {\rm diag}\,( m_1,\, m_2,\,...,m_N\,)\ ,\quad m_i\neq 0\ ,
\ee
where $m_i\neq m_j$ if $i\neq j$.
This results in $N$ first order equations
\be
\cD \left( \tfrac{\lambda s}{m_i} \right) \Phi^{\,i}_{(s)} =0 \ , \qquad   i=1,...,N\ , \qquad \mbox{(no sum)}\ .
\ee
Since all $m_i$ are assumed to be distinct, the solutions of these equations are those  of
\be
\cD \left( \tfrac{\lambda s}{m_1} \right)\, \cD \left( \tfrac{\lambda s}{m_2} \right) \cdots \cD \left( \tfrac{\lambda s}{m_r}\right) \Phi_{(s)} =0\ .
\label{dyneq1}
\ee
The case of $N=2$ and $m_1=-m_2 \equiv m$ gives  \eq{fe} discussed above.
Taking $N=2$ and $m_1 \ne - m_2$ gives a higher spin generalization of
generalized massive gravity.

Finally, if the mass matrix cannot be diagonalized, it can be brought into a direct sum of Jordan blocks, 
each given by $r\times r$ blocks of the form
\be
\mu_{(r)}= \begin{pmatrix}   m & \lambda   &  & &     \\
                                               & m  &\lambda &   &   \\
                                              & &  \ddots & \ddots  &   &   \\
                                             & &  &  m   & \lambda \\
                                             & & & &  m
                  \end{pmatrix}
\ee
In this block, equations \eq{ne2} take the form
\bea
\cD (\tfrac{\lambda s}{m} )\  \Phi^{\, i}_{(s)}  &=& -\tfrac{s}{\lambda} \Phi^{i+1}_{(s)}\ , \quad  i=1,...,r-1\ ,
\w2
\cD(\tfrac{\lambda s}{m}) \Phi^{\,r}_{(s)} &=& 0\ .
\eea
Eliminating  the fields $\Phi^n_{(s)}$ in terms of $\Phi^{n-1}_{(s)}$ successively, and 
denoting $\Phi^{1}_{(s)} \equiv \Phi_{(s)}$,  one finds
\be
\left[ \cD(\tfrac{\lambda s}{m}) \right]^r\ \Phi_{(s)} =0\ .
\label{dyneq2}
\ee
These are higher spin generalizations of 3D critical massive gravities.
In addition to the standard spin-s solution with lowest AdS energy $E_0=1+\frac{m}{\lambda s}$, they have p-fold  logarithmic
solutions for $p=1,...,r-1$.
For $s=2$ they have been studied extensively, see for example, \cite{Kogan:1999bn, Ghezelbash:1998rj} 
where their holographic duals have been shown to be
Logarithmic CFTs.
The issue of unitarity of these models has also been investigated.
In particular, if $r$ is odd, it has been argued in \cite{Bergshoeff:2012sc} by means of a toy
model for free scalar fields that there exists a unitary truncation of the theory that maintains some of
the logarithmic modes.

\section{Introduction of gauge potentials}
\label{sec:field}

It is a general property of unfolded equations that the gauge invariant local degrees of
freedom are fully encoded in the sector of zero forms.
The description of their interactions, however, requires the introduction of gauge potentials.
Sometime ago, Prokushkin and Vasiliev \cite{Prokushkin:1998bq}
constructed  fully nonlinear higher spin gravity theory in three dimensions in terms of a master zero-form
field and a master one-form gauge field. For generic values of a deformation parameter given by an expectation
value, these equations  describe only massive scalar degrees of freedom,
while gauge fields with spin $s\ge 1$ are topological.  At critical values of the PV deformation parameter, one recovers \eq{L2} in
the zero-form sector.  At present, it is not known how to include the one-form in the linearization of the critical PV system, as
we shall discuss further in the conclusions. Moreover, the deformation of the full PV  equations to include our deformation
parameter $\mu$ is not known.

Our aim here is to introduce  gauge potentials for the $\mu$-deformed system, in a fashion  that  utilizes the unfolding techniques.
As we are working at the linearized level without an underlying nonabelian higher spin algebra,  we will study one spin at a time rather
than packaging them all in a one-form master field.

While the unfolded formulation is indispensable for the description of higher spin gravities with local
degrees of freedom,  there is no unique way of introducing the  required gauge potentials \cite{Vasiliev:1988xc,Skvortsov:2008vs,Boulanger:2008up,Boulanger:2012mq}\footnote{A similar issue was discussed recently in the context of $AdS_2$ in \cite{Alkalaev:2013fsa}, where the gauge
potentials introduced therein transform like maximal depth partially-massless fields. The Weyl module of that
theory is similar to the $AdS_3$ $\mu$-deformed  Weyl modules we constructed here, so 
one can expect that gauge modules of partially-massless fields can be consistently used for
$3d$ $\mu$-deformed systems as well.}. 
Indeed, as we shall see,
there are two natural ways to let the zero-form source  properly defined linearized higher spin  curvatures. In one approach,
we shall use trace-unconstrained metric-like potentials which lead to  primary field equations generalizing  the
results of \cite{Bergshoeff:2009tb}  in Minkowski space, and for special values of the spin, to arbitrary spins in $AdS_3$.
This approach  leads to an important single trace condition on the dual of the generalized  Riemann tensor
(equivalent to double-trace condition on the generalized Riemann tensor itself). In the second approach,
we shall  employ Fronsdal-type trace-constrained  gauge potentials and the above mentioned trace-conditions appear
naturally here as well. Thus, in both of these approaches, the unfolded formalism that we are  using makes it far more convenient
to demonstrate the decoupling of redundant gauge degrees of freedom contained in  the  gauge potentials.

\subsection{Trace-unconstrained gauge potentials}

We introduce a set of one-forms
\begin{equation}
w= \{ \omega_{m(s-1),n(t)}\}\ , \quad t=0, 1,\ldots, s-1\ ,
\end{equation}
taking their values in two-row-shaped Young tableaux of  $gl(3)\,$, with the first
row of length $s-1$ and the second of length $t\,$. The structure group is ${so}(1,2)$. We use $m,n,...$ to
denote indices on the  trace-unconstrained objects and we use $h_\mu^m$ to convert between flat and curved
indices. The unfolded equations that we propose read
\be
 \nabla \omega_{m(s-1),n(t)} -
  h^p \,\omega_{m(s-1),n(t)p}
  -\rho_t\,\left(h_n\, \omega_{m(s-1),n(t-1)}-\tfrac{s-1}{s-t}\;h_m \,\omega_{m(s-2)n,n(t-1)}\right)=0\ ,
\label{gs1}
\ee
for ${t < s-1}$ and
\bea
&& \nabla\omega_{m(s-1),n(s-1)} -\rho_{s-1}\Big( h_n \,\omega_{m(s-1),n(s-2)} -  {(s-1)} \,h_m\,\omega_{m(s-2)n,n(s-2)}\Big)
\nn\\
&& \qquad \qquad  = h^p \,h^q \;\epsilon_{pq}{}^{c_{s}}  \epsilon_{m_{1}n_{1}}{}^{c_{1}}\dots
  \epsilon_{m_{s-1}n_{s-1}}{}^{c_{s-1}} \,\Phi_{c_{1}\ldots c_{s}}\ ,
  \label{gs2}
\eea
for $t=s-1$ and where $\Phi_{c(s)}$ is the  primary zero-form discussed in Section 2,
obeying \eq{pp}. In establishing  the integrability of the above equations  we
use \eq{pp} and it is required that
\begin{eqnarray}
\rho_t = \lambda^2 t(s-t)\ .
\label{rho1}
\end{eqnarray}
Equations \eq{gs1} and \eq{gs2} possess the following gauge invariances:
\bea
{t < s-1}: \quad  \delta_{\epsilon} \omega_{m(s-1),n(t)} & = &
  \nabla \epsilon_{m(s-1),n(t)} - h^p \,\epsilon_{m(s-1),n(t)p}
  \label{gtransfo1}\\
&& - \rho_t\,\left(h_n\, \epsilon_{m(s-1),n(t-1)} - \tfrac{s-1}{s-t}\; h_m \,\epsilon_{m(s-2)n,n(t-1)}\right)\ ,
\nn\\
{t = s-1}: \;\delta_{\epsilon} \omega_{m(s-1),n(s-1)} &=  & \nabla \epsilon_{m(s-1),n(s-1)}
\label{gtransfo2}\\
&&  - \rho_{s-1}\left(h_n \,\epsilon_{m(s-1),n(s-2)} -  {(s-1)} \,h_m\,\epsilon_{m(s-2)n,n(s-2)}\right)\ ,
\nn
\eea
where the parameters $\{ \epsilon_{m(s-1),n(t)}\}_{t=0,1,\ldots, s-1}$ are zero-forms
that possess the same  $gl(3)$ irreducible symmetries as their corresponding one-form connections.
The one-forms $\omega_{m(s-1)}$ and $\omega_{m(s-1), n}$ provide possible higher spin  analogs of the gravitational
dreibein and spin connection,  respectively, while $\omega_{m(s-1),n(t)}$ for $t>1$ are auxiliary fields.
From  \eqref{gs1} and  \eqref{gs2} it follows that  all the components of the connections can either be
set to zero by using the algebraic gauge transformations  contained in \eqref{gtransfo1} and
\eqref{gtransfo2}, or be expressed as derivatives of the unconstrained metric-like field
\begin{equation}
h_{\mu_{1}\ldots \mu_{s}}:= s\, h_{(\mu_2}^{m_2}\cdots h_{\mu_s}^{m_s}\, \omega_{ \mu_1) \vert m_2 \ldots m_s}\ .
\end{equation}
One can thus  write \eq{gs2} as
\be
R_{p_1q_1\vert \ldots |p_s q_s} =   (-\tfrac12)^s\, \epsilon_{p_1q_1}{}^{m_1} \ldots
\epsilon_{p_s q_s}{}^{m_s}\, \Phi_{m_1\ldots m_s}\ ,
\ee
where the linearized generalized Riemann tensor is defined by
\be
R_{p_{1}q_{1}\vert \ldots |p_{s}q_{s}}  :=  \delta_{[p_1}^{p_1'} \delta_{q_1]}^{q_1'} \cdots
\delta_{[p_s}^{p_s'} \delta_{q_s]}^{q_s'}\,   \nabla_{p_1'}\ldots\nabla_{p_s'} h_{q_1'\ldots q_s'} + {\cal O}(\lambda^{2})\ ,
\label{RDF}
\ee
where the order $\lambda^2$ terms are determined by the requirement of invariance under the abelian gauge transformations
$\delta h_{\mu_1\ldots \mu_s}= s\nabla_{(\mu_1} \epsilon_{\mu_2\ldots \mu_s)}$.  Its  precise form can be found
in \cite{Manvelyan:2007hv}.  It is well-known that such a tensor is unique up to Hodge dualisations and it is
a generalisation of the  de Wit--Freedman curvature  \cite{deWit:1979pe} to AdS space.  The Hodge dual of this curvature, which we
shall denote by ${\widetilde R}_{m_{1}\ldots m_{s}}$ is
defined as
%
\begin{equation}
{\widetilde R}_{m_{1}\ldots m_{s}} =
\epsilon_{m_1}{}^{p_1q_1}\ldots \epsilon_{m_s}{}^{p_sq_s}\,R_{p_{1}q_{1}\vert \ldots |p_{s}q_{s}}\ .
\label{3dv}
\end{equation}
Thus, equation  \eq{gs2} takes the simple form
 \be
{\widetilde R}_{m_{1}\ldots m_{s}} = \Phi_{m_1...m_s}\ ,
 \label{tmg2}
 \ee
which imposes the tracelessness condition
\be
\label{einsttraceless}
{\widetilde R}'_{\mu_{1}\ldots \mu_{s-2}}  = 0\ .
\ee
Combining  \eq{tmg2} with \eq{tmg}  gives
\bea
&& {\widetilde R}_{\mu_1...\mu_s}
+ \frac{s}{\mu} \,\varepsilon_{(\mu_1|}{}^{ab} \nabla_a {\widetilde R}_{b|\mu_2...\mu_s)} =0\ .
\label{g1}
\label{gextra}
\eea
Thus, the full set of linearized field equations are given by \eq{g1} and \eq{einsttraceless}.
The spectral analysis has been performed in 3D Minkowski space in \cite{Bergshoeff:2009tb,Bergshoeff:2011pm} 
for $s=3$ and $s=4$, where thus the equations are of order 4 and 5, respectively.
Indeed, they propagate a single degree of freedom, which
is consistent with the fact that the gauge invariant primary
zero-form obeys a first-order equation.

In the case of several families of zero-forms discussed in Section 3, the unfolded system
can be extended with gauge potentials in an analogous way. Namely, in (\ref{gs2})
$\Phi_{c_1\dots c_s}$ has to be replaced with the primary tensor of the
first family of zero-forms $\Phi^{1}_{c_1\dots c_s}$. This leads to 
 \be
{\widetilde R}_{m_{1}\ldots m_{s}} = \Phi^{1}_{m_1...m_s}\ .
 \label{nmg2}
 \ee
The primary Weyl tensor $\Phi^{1}_{m_1...m_s}$ is in turn subjected to equations of motion
such as (\ref{dyneq1}) or (\ref{dyneq2}). Together with (\ref{nmg2}) they give
the equations of motion in terms of gauge potentials. For example, for the 
higher spin extension of new massive gravity from (\ref{nmg}) one gets 
\bea
\Box{\widetilde R}_{\nu(s)} - \left( \frac{m^2}{s^2} - \lambda^2 (s+1)\right)
{\widetilde R}_{\nu(s)} &=& 0\ .
\label{nmg3}
\eea
Let us note once again that (\ref{nmg2}) implies the tracelessness condition 
(\ref{einsttraceless}).

Another possible extension of the presented construction is to consider 
several families of one-forms. Let them be enumerated by an index $\mu = 1, \dots, M$. Then
instead of (\ref{gs1}) and (\ref{gs2}) one has
\be
 \nabla \omega^{(\mu)}_{m(s-1),n(t)} -
  h^p \,\omega^{{(\mu)}}_{m(s-1),n(t)p}
  -\rho_t\,\left(h_n\, \omega^{{(\mu)}}_{m(s-1),n(t-1)}
  -\tfrac{s-1}{s-t}\;h_m \,\omega^{{(\mu)}}_{m(s-2)n,n(t-1)}\right)=0\ ,
\label{gsmult1}
\ee
\bea
&& \nabla\omega^{(\mu)}_{m(s-1),n(s-1)} -\rho_{s-1}\Big( h_n \,\omega^{(\mu)}_{m(s-1),n(s-2)} 
-  {(s-1)} \,h_m\,\omega^{(\mu)}_{m(s-2)n,n(s-2)}\Big)
\nn\\
&& \qquad \qquad  = h^p \,h^q \;\epsilon_{pq}{}^{c_{s}}  \epsilon_{m_{1}n_{1}}{}^{c_{1}}\dots
  \epsilon_{m_{s-1}n_{s-1}}{}^{c_{s-1}} B^\mu{}_i \,\Phi^{i}_{c_{1}\ldots c_{s}}\ ,
  \label{gsmult2}
\eea
where $B^\mu{}_i$ is some constant matrix intertwining zero- and one-forms, and  $\Phi^i_{c(s)}$ are the  
primary zero-forms discussed in Section 3, obeying \eq{pp2}. Even though the modifications of the 
gauge sector of the unfolded equations do not affect the dynamical content of the theory, they may 
turn out to be necessary to construct interactions. In particular, the $M=N$ case may play a special role since 
it might as the linearisation of the critical PV system for matrix-valued fields \cite{Prokushkin:1998bq}.  

The following additional remarks are in order.  Firstly, the condition
\eq{einsttraceless} simplifies \eq{RDF} so that it can be written simply as
\be
{\widetilde R}_{m_{1}\ldots m_{s}}(h) =
\epsilon_{m_1}{}^{p_1q_1}\ldots \epsilon_{m_s}{}^{p_sq_s}\, 
\nabla_{(p_{1}}\ldots\nabla_{p_{s})}h_{q_{1}\ldots q_{s}}  -{\rm traces}\ .
\label{RDF2}
\ee
Second, working with metric-like gauge fields in $D\geqslant 4$,
the linearized spin$-s$ Riemann tensor is given by $s$ curls of the 
gauge potential (just like in 3D).
However, unlike 3D, it decomposes into a traceless part
given by the spin-$s$ Weyl tensor (which has no 3D analog)
and a trace part given by $s-2$ curls of the Fronsdal tensor.
Hence, equating Riemann and Weyl tensors in $D\geqslant 4$,
yields the Fronsdal equation, putting the theory completely on-shell.
Moreover, the integrability of this curvature
constraint implies the complete unfolding system of the
Weyl zero-form.
In $3D$, however, the Weyl tensor vanishes identically,
and instead we equate the rank-$s$ primary zero-form to
the Hodge-dualised Riemann tensor as in \eq{tmg2}.
Tracing this constraint sets to zero the doubly traced
generalized Riemann tensor, which removes an unphysical
spin-$(s-2)$ mode. The equation of motion \eq{pp} then supplies the 
dynamical equation of motion  for the trace-free spin-s field.

Finally, the consistency of the combined system of equations   \eq{gsmult2} and \eq{pp2} can be 
seen as follows. The unfolded equation for the primary Weyl tensors $\Phi^i_{(s)}$ compatible
with the curl of  \eq{gsmult2} is
\bea
B^\mu{}_i \left[ \nabla \Phi^i_{a(s)}-  h^b \left( \Phi^i_{b a(s)}+\epsilon_{ba}{}^c \, \Psi^i_{a(s-1)c}\right)\right] = 0\ ,
\label{pp3}
\eea
where $\Phi^i_{b a(s)}$ and $\Psi^i_{a(s-1)c}$ are symmetric and traceless but otherwise arbitrary tensors. 
Thus,  setting $\Psi^i_{a(s)} = \mu^i{}_j \Phi^j_{a(s)}/(s+1)$ where $\mu^i{}_j$ are arbitrary real  constants, 
we observe that  this equation is satisfied for any $B^\mu{}_i$  in view of \eq{pp2}, which we have already 
established in Section 3 to describe a consistent set of equations.
%

\subsection{Trace constrained gauge potentials}

Another natural possibility is to use a set of {\it traceless} one-forms
\begin{equation}
v = \{  v_{a(s-1),b(t)}\}\ , \quad t=0, 1\ ,
\end{equation}
taking values in one and two-row-shaped tensors of $so(1,2)\,$. Focusing on the case of a 
single zero-form tower ($N=1)$,
the unfolded equations then read
\begin{eqnarray}
  \label{gs4}
\,\nabla v_{a(s-1)} &=  & h^b \,v_{a(s-1),b}\ ,  \\
\label{gs5}
 \,\nabla v_{a(s-1),b} &=  &
  h^c \, \left( h_b \,\Phi_{a(s-1)c} - h_a\, \Phi_{a(s-2)bc}\right) +
  \\ \nonumber
& & \tau\left(h_b \, v_{a(s-1)} - h_a \, v_{a(s-2)b}+\tfrac{s-2}{s-1}\left[h_c \,v^{c}{}_{b a(s-3)}
 \,\eta_{aa} - h_c \, v^c{}_{a(s-2)}\,\eta_{ab}\right]   \right)\ ,
\end{eqnarray}
where $\Phi_{c(s)}$ is the  primary zero-form  obeying \eq{pp}. The Cartan-Frobenius integrability 
of these equations requires
\begin{eqnarray}
\tau=(s-1)^2\ .
\end{eqnarray}
Equation \eqref{gs4} is the zero-torsion condition for the unfolded
Lopatin--Vasiliev system \cite{Lopatin:1987hz} in $AdS_{3}\,$.  Using this condition to eliminate $v_{a(s-1),b}$,  
equation \eq{gs5} takes the form
\be
F_{\mu_1...\mu_s} =\Phi_{\mu_1...\mu_s}\ ,
\label{ff}
\ee
with Fronsdal tensor defined as \cite{Fronsdal:1978rb}
\begin{equation}
\label{Fronsdaltens}
{ F}_{\mu_1...\mu_s}  = \Big[\Box -s(s-3)\Big] \varphi_{\mu_1...\mu_s}
-s\nabla_{(\mu_1} \nabla^\nu \varphi_{\mu_2...\mu_s)\nu}
+\frac{s(s-1)}{2} \left[ \nabla_{(\mu_1}\nabla_{\mu_2}
- 2 g_{(\mu_1\mu_2} \right] \varphi'_{\mu_3...\mu_s)}\ ,
\end{equation}
where we have set $\lambda=1$ and $\varphi_{a(s)}$ is the  the doubly  traceless gauge field
\be
\varphi_{a(s)}:=s\, v_{a | a(s-1)}\ ,\qquad \varphi''_{a(s-4)}=0\ .
\label{vf}
\ee
Equation \eq{ff}  imposes the tracelessness condition on the Fronsdal tensor. Thus, 
combining \eq{ff} with \eq{tmg} we end up with the dynamical field equations
\begin{eqnarray}
&& { F}_{\mu_1...\mu_s}
+ \frac{s}{\mu} \,\varepsilon_{(\mu_1|}{}^{ab} \nabla_a {F}_{b|\mu_2...\mu_s)}  =0\ ,
\label{f1}\\
&& {F}'_{\mu_1...\mu_{s-2}}  =0\ .
\label{extra}
\end{eqnarray}
For $s=1$, equation \eqref{f1} describes a topologically
massive photon.
For $s=2$, equations \eqref{f1} and \eqref{extra}
describe topologically massive gravity, and for
$s \geqslant 3$ they provide a generalization
thereof to higher spins.

In comparison, in $D\geqslant 4$, the formulations in terms
of trace constrained potentials, the corresponding set of linearized curvatures
consists of the Fronsdal tensor, generalized torsions and the
generalized spin-$s$ Weyl tensor.
The linearized curvature constraints set the Fronsdal tensor,
and not just its $(s-2)$th curl, and the generalized torsions to
zero, leaving the spin-$s$ Weyl tensor given by the traceless
projection of $s$ curls of the trace constrained
gauge potential.
The latter curvature constraint then induces the complete
unfolding system of the Weyl zero-form.
In $3D$, however, as we have just seen, the use of trace constrained potentials
leads to an identification of the primary zero-form
with the (two-derivative) Fronsdal tensor, and the
fields go on-shell as the result of the first level
of equations in the Weyl zero-form module, which
we have imposed by hand, as noted earlier.

Finally, in order to exhibit the extra boundary states that
arise in the gauge potentials in $AdS_3$, we impose the de
Donder gauge and removing the trace mode by using
\eq{extra} and on-shell gauge transformations.
From \eq{f1} and \eq{extra} we thus find
\bea
&& \left[\Box -s(s-3)\right] \left( \varphi_{\mu_1...\mu_s}
+\frac{s}{\mu} \varepsilon_{(\mu_1|}{}^{ab} \nabla_a \varphi_{b|\mu_2...\mu_s)}\right)=0\ ,
\nn\\
&& \varphi'_{a(s-2)}=0\ ,\qquad  \nabla^b \varphi_{ba(s-1)}=0\ .
\eea
Expanding $\varphi$ in lowest-energy UIRs, the resulting lowest weights
are
\begin{equation}
(E_0,s_0):\qquad (s,s)\ ,\qquad (s,-s)\ ,\qquad (1+\frac{\mu}{s}, s)\ .
\end{equation}
The first two states are boundary singleton states that
arise in the gauge potential but not in the primary zero-form, which
only contains the last state.
If $\mu=s(s-1)$ then the latter mode disappears but a logarithmic mode
appears due to degeneracy with the singletons in the gauge field,
see \emph{e.g.} the discussions in the case of critical gravity in
in \cite{Bergshoeff:2010iy}.
In the $\mu \to 0$ limit, the massive spin-$s$ state turns into a
state with lowest weight $(1,s)$, which is the spin-$s$ analog of the $(1,2)$
state that arises in conformal Chern--Simons gravity as a partially massless state \cite{Afshar:2011qw}.

\section{Conclusions}

We have constructed unfolded formulations providing
a higher spin generalization in flat spacetime as well
as AdS$_3$ of linearized theories 
of massive gravity with higher derivatives such as 
topologically massive gravity, general massive gravity
and their critical versions.
We have also analyzed their spectrum and exhibited unitary
as well as interesting non-unitary representations, including
higher order singletons and finite-dimensional representations
of $so(2,2)$ arising at critical points. 
The latter leads to indecomposable structures that are similar
to those in the linearized PV model found in \cite{Barabanshchikov:1996mc,Prokushkin:1998bq}, 
though there are two quite different parameters that govern the critical limits.
We have also glued two different types of gauge potentials to the zero-form
sector.
Although these obey quite different equations of motion, in one case
being of third order in derivatives and in the other case being
of order $s+1$ in derivatives, the unfolded formulation guarantees 
that they propagate the same local bulk degrees of freedom, though
boundary states may differ, which we leave for future studies.

It seems natural to conjecture that the ultimate construction 
of a fully nonlinear massive higher spin gravity in 3D will 
require elements of the existing Prokushin--Vasiliev higher 
spin theory \cite{Prokushkin:1998bq}, which in its present 
and generic form describes massive scalars coupled to 
nonpropagating topological higher spin fields.  
Indeed one of the main motivations for the present paper has  
been to provide a convenient framework for such a construction 
by first tackling the desired linearized field equations. 
At that level, we have seen that our unfolded equations for 
the zero-form module correspond to deformation of the 
linearized PV system at its critical point
by a parity breaking mass parameter $\mu$.  
The full $\mu$-deformed PV theory, however, 
remains to be constructed. 
It may involve a generalization of the PV system, 
or possibly, and more simply, the expansion around a new vacuum solution  
such that the resulting linearized field equations accommodate
our deformation parameter $\mu$.  
Possibly, progress can also be made by a closer examination 
of the curvature constraint in the critical PV system, 
and seeking a finite result for the source term that 
depends on the zero-form master field and that naively 
diverges at the critical point.

\vspace{0.5cm}

\subsection*{Acknowledgments}

We thank Mauricio Valenzuela, Fabien Buisseret and Nicol\`{o} Colombo
for their collaboration at the early stages  of this paper. 
It is also a pleasure to thank Thomas Basile, Eric Bergshoeff, Daniel Grumiller,
Kostas Siampos, Philippe Spindel, Misha Vasiliev and Yihao Yin for useful discussions.
The work of N.B. was partially supported by the contract
``Actions de Recherche concert\'ees -- Communaut\'e fran\c{c}aise de  Belgique''
AUWB-2010-10/15-UMONS-1, that of E.S. by NSF grant PHY-1214344 and that of
D.P. by the DFG grant HO 4261/2-1 ``Generalized dualities relating
gravitational theories in 4-dimensional Anti-de Sitter space with 3-dimensional conformal field theories''.
N.B. is a Research Associate of the Fonds de la Recherche Scientifique\,-FNRS (Belgium).   

\begin{appendix}

\section{Conventions and  decomposition of $so(2,2)$ }
\label{sec:A}

\paragraph{Conventions.}

The anti-de Sitter AdS$_{d+1}$ algebra $so(2,d)$ is presented by
hermitian generators $M_{AB}$ obeying the commutation relations
\begin{eqnarray}
[M_{AB},M_{CD}] = {\rm i}\, \eta_{BC}M_{AD} - {\rm i} \,\eta_{AC}M_{BD}
   -  {\rm i}\, \eta_{BD}M_{AC} + {\rm i} \,\eta_{AD}M_{BC}\ ,
\end{eqnarray}
where $(\eta_{AB}) =$ diag$(-,-,+,+)\,$ and the $so(2,2)$ vector index
$A = 0',0,1,2\,$, also decomposed as  $A = 0',a\,$ in terms of the Lorentz $so(1,2)\subset so(2,2)$ index $a=0,1,2\,$.
Hence, $(\eta_{ab}) =$ diag$(-,+, +)\,$. The AdS$_3$ transvection operators are defined by
\be
P_a := \lambda\,M_{0' a}\ ,
\ee
so that the AdS$_3$ algebra is presented by
\begin{eqnarray}
[M_{ab},M_{cd}] = 2{\rm i}\, \eta_{c[b}M_{a]d} - 2{\rm i}\,\eta_{d[b}M_{a]c}  \;, \quad
[P_a,P_b] = {\rm i} \,\lambda^2\, M_{ab}\;, \quad
[M_{ab},P_c] = 2{\rm i} \,\eta_{c[b}P_{a]} \ .
\end{eqnarray}
The action of the Lorentz generator in the vector representation is
\begin{eqnarray}
\rho_V(M_{ab})V_c  = - 2{\rm i} \,\eta_{c[b}V_{a]}\;
\label{vectorrepLor}
\end{eqnarray}
and the Lorentz-covariant differential operator
is
\begin{eqnarray}
\nabla := {\rm d} - \tfrac{{\rm i}}{2}\, \omega^{ab}\rho(M_{ab})\ .
\end{eqnarray}
The AdS$_3$ differential $\cal D$ is defined by
\begin{eqnarray}
{\cal D} := {\nabla} - {\rm i}\,  \,h^a \rho(P_a) =
{\rm d} + \tfrac{1}{2{\rm i}}\, \Omega^{AB}\rho(M_{AB}) = {\rm d} + \Omega\ ,
\end{eqnarray}
where the vielbein one-forms $h^a := \lambda^{-1}\Omega^{0' a}$ are required to have invertible
components.
Explicitly, the flatness condition on the connection $\cal D$ gives
\begin{eqnarray}
\nabla^2 \equiv \tfrac{1}{2{\rm i}}\, R^{ab} \rho(M_{ab}) = \tfrac{\rm i}{2}\, \lambda^2 \,h^a h^b\, \rho(M_{ab})\ ,
\end{eqnarray}
so that, in particular,
\begin{eqnarray}
\nabla^2 V^a = -  \lambda^2 h^a h_b\, V^b\ .
\end{eqnarray}
In the metric formulation where ${g}_{\mu\nu}$ denote the components
of the AdS$_3$ metric, this implies that the components of the
Riemann tensor are given  by
\begin{eqnarray}
R_{\mu\nu\alpha\beta} = - \lambda^2
({g}_{\mu\alpha}{g}_{\nu\beta}- {g}_{\nu\alpha}{g}_{\mu\beta})\ .
\end{eqnarray}
%
\paragraph{Decomposition.} 
%
The ${so}(2,2)$ algebra has two irreducible subalgebras ${so}(1,2)_{(\pm)}$ of ${so}(2,2)$
generated by $J^{(\varepsilon)}_a$ where $\varepsilon=\pm$. These subalgebras are
defined as
\bea
& J^{(\varepsilon)}_a := \tfrac{1}{2}\, (M_a + \frac{\varepsilon}{\lambda} P_a\,)\ ,\quad
M^{a}~:=~\tfrac{1}{2}\, \epsilon^{abc} M_{bc}\ ,\quad \epsilon^{012} = 1\ ,&
\label{SpinEnergy}\\
& [ J^{(\varepsilon)}_a,J^{(\varepsilon)}_b ] = (-{\rm i})\,\epsilon_{abc} \,\eta^{cd} J^{(\varepsilon)}_d\ ,\quad
[J^{(+)}_a,J^{(-)}_b]~=~0\ .&
\eea
In the compact basis the algebra ${so}(2,2)$  takes the form
\begin{equation}
J^{(\varepsilon)}_{\pm}~:=~ J^{(\varepsilon)}_1 \pm {\rm i} J^{(\varepsilon)}_2\ ,\quad
[J^{(\varepsilon)}_0,J^{(\varepsilon)}_\pm]~=~\pm J^{(\varepsilon)}_\pm\ ,\quad
[J^{(\varepsilon)}_\mp, J^{(\varepsilon)}_\pm] ~=~\pm 2J^{(\varepsilon)}_0\ .
\label{a12}
\end{equation}
We normalize the quadratic Casimirs as follows  ($A,B=0',a$ with $a= 0, j\,$ and $j=1,2$)
\bea
C_2[{so}(2,2)] &=& \frac12 M^{AB} M_{AB}\ ,
\qquad C_2[{so}(1,2)_{\rm Lor}] = \frac12 M^{ab} M_{ab}\ ,
\label{C1}\w2
C_2[{so}(1,2)_{(\varepsilon)}] &=& \eta^{ab} J_a^{(\varepsilon)} J_b^{(\varepsilon)}\ ,
\label{C2}
\eea
It follows that
\bea
C_2[{so}(2,2)] &=& C_2[{so}(1,2)_{\rm Lor}]  - P_a P^a\ ,
\label{C3}\w2
C_2[{so}(1,2)_+] + C_2[{so}(1,2)_-] &=& -\frac12 C_2[{so}(2,2)]\ ,
\label{C4}\w2
C_2[{so}(1,2)_+] - C_2[{so}(1,2)_-]  &=& M^a P_a\ .
\label{C5}
\eea
From the last two expressions we get
\be
C_2[{so}(1,2)_{(\varepsilon)}] =  -\frac12 C_2[{so}(2,2)] + \frac{\varepsilon}{2} M^a P_a\ ,
\label{C6}
\ee
which plays an important role in section 3.

In constructing the representations of the algebra ${so}(2,2)$, it is useful to split its
generators $M_{AB}\,$ into
\be
 M_{12}\ , \quad L^\pm_j := M_{0j}\mp {\rm i}M_{0'j}\ , \quad
E := \lambda\, M_{0'0}\equiv P_0\ , \quad j=1,2\ ,
\label{ke}
\ee
which obey the algebra
\be
[E,L^\pm_j] = \pm \lambda\, L^\pm_j\ , \qquad [L^+_i,L^-_j] =  2{\rm i}M_{ij} - 2\lambda^{-1}\, \delta_{ij}E\ .
\ee
The operators of spin and energy are given by
\begin{eqnarray}
E := P_{0} = J_{0}^{(+)} - J_{0}^{(-)}\;,\qquad S := M_{12} = - ( J_{0}^{(+)} + J_{0}^{(-)} )\ .
\label{spinandenergy}
\end{eqnarray}
In a lowest-weight irreducible representation of  $so(2,2)\,$ with vacuum
$\vert E_0,{s}\rangle\,$, one has
\begin{eqnarray}
L^-_i \vert E_0,{s_0}\rangle = 0 = (\lambda^{-1}\,E-E_0)\vert E_0,{s_0}\rangle\ ,
\end{eqnarray}
so that
\begin{eqnarray}
C_2[so(2,2)] \vert E_0,{s_0}\rangle &=& \left( C_2[so(2)] - \delta^{ij} L^+_i L^-_j
- E_0 (- E_0 + 2) \right)  \vert E_0,{s_0}\rangle\;,
\nn\\
&=& \left( - E_0 [- E_0 + 2] + C_2[so(2)]\, \right)\,  \vert E_0,{s_0}\rangle\ ,
\label{A14}
\end{eqnarray}
where $C_2[so(2)]:=\tfrac{1}{2}\,M^{ij}M_{ij}\,=S^{2}\,$.

\section{Oscillator formulation of the zero form module}
\label{sec:B}

We define the one-forms
\begin{align*}
\boldsymbol{\sigma}^-  & := (- {\rm i})\, E^-\ ,   &  \boldsymbol{\sigma}^0  &:= (- {\rm i}) \,\frac{M(N)}{N}\, {\cal E} \ , &
\boldsymbol{\sigma}^+ & := (- {\rm i})\, \frac{\Lambda(N)}{N} \,\pi \,E^+\ ,
\\
E^-  & := h^a {\alpha}_a\ ,   & {\cal E}  & := h^a \epsilon_{ab}{}^c \bar{\alpha}^b \alpha_c\ , &  E^+ & := h_a \bar{\alpha}^a\ ,
\end{align*}
where $\Lambda(N)$ and $M(N)$ are functions of the number operator. $N := \bar{\alpha}^a \alpha_a $ and
\be
\pi := \left(\mathbb{1} - \frac{1}{(2N-1)}\,\bar{T}T\right) \ ,    \qquad  T := \eta^{ab}\alpha_a\alpha_b\ , \quad
\bar{T} := \eta_{ab}\, \bar\alpha^a\bar \alpha^b\ .
\ee
Introducing $H_a := \epsilon_{abc}h^bh^c$, one has
\begin{eqnarray}
\left\{ {\cal E} ,  \pi \right\} = 0 \, \quad
\left\{ {\cal E} ,  {\cal E} \right\} = -2 E^+E^- \ , \quad
\left\{ {\cal E} ,  E^- \right\} = - H^a \alpha_a  \ , \quad
\left\{ {\cal E} ,  E^+ \right\} = - H_a \bar\alpha^a  \ .
\label{littlelemma}
\end{eqnarray}

We define the  zero-form ${so}(2,2)$-module $\widetilde{\cal T}_{(0)}$
via the representation $\tilde{\rho}(P_a)$ of the $AdS_3$ transvection generators,
the Lorentz generators acting tensorially, \emph{i.e.} according to \eqref{vectorrepLor}.
Therefore, when acting in $\widetilde{\cal T}\,$, the AdS$_{3}$ connection ${\cal D}$ is
\begin{eqnarray}
{\cal D} := \nabla - {\rm i} \, h^a \tilde{\rho}(P_a) =
\nabla - {\rm i} \,(\boldsymbol{\sigma}^-
            +\boldsymbol{\sigma}^0 +\boldsymbol{\sigma}^+) \ ,
\label{calD}
\end{eqnarray}
and the unfolded, linear equations \eqref{unfeq}  for the propagation of fields in $AdS_3$ are given by
\begin{eqnarray}
{\cal D} \vert \Phi^{(s)} \rangle = 0\;.
\label{UnfoldedEq}
\end{eqnarray}
In comparing this equation with \eq{unfeq}, we note that  $M(n) := \mu_n$ and $\Lambda(n) := \lambda_n$.
The Cartan-Frobenius integrability of these equations guaranties that
the set of zero-forms in $\vert \Phi^{(s)} \rangle$ forms a representation of $so(2,2)\,$,
where each of the zero-form field $\Phi_{a(n)}$ in $\vert \Phi^{(s)} \rangle$
transforms in the totally-symmetric
rank-$n$ representation of the Lorentz subalgebra $so(1,2)_{\rm Lor}\subset so(2,2)\,$.

Decomposing the integrability condition into linearly independent pieces
gives the five \emph{a priori} independent conditions
\begin{eqnarray}
& \left( \boldsymbol{\sigma}^- \right){}^2 = 0\ , \qquad
\left\{ \boldsymbol{\sigma}^- ,  \boldsymbol{\sigma}^0 \right\}  = 0\ , \qquad
\nabla^{2} + \left\{ \boldsymbol{\sigma}^- ,  \boldsymbol{\sigma}^+ \right\}
+ \left( \boldsymbol{\sigma}^0 \right){}^2   = 0\ , \qquad &
\\
& \left\{ \boldsymbol{\sigma}^0 ,  \boldsymbol{\sigma}^+ \right\}
= 0\ , \quad \left( \boldsymbol{\sigma}^+ \right){}^2 = 0\ . \qquad &
\end{eqnarray}
The first and last consistency equations above are identically satisfied, as can be seen
from the definition of  $\boldsymbol{\sigma}^{\pm}\,$.
In order to evaluate the left-hand sides of the remaining three equations, we
use \eqref{littlelemma}. This yields
\begin{eqnarray}
\left\{ \boldsymbol{\sigma}^- ,  \boldsymbol{\sigma}^0 \right\} =
- \tfrac{1}{2}\,H^a \alpha_a \left[  M(N-1) - \frac{(N +1)}{N}\, M(N)\right]\ .
\end{eqnarray}
The consistency equation  $\left\{ \boldsymbol{\sigma}^- ,  \boldsymbol{\sigma}^0 \right\} = 0$
provides us with a recursion relation
\be
n\,\mu_{n-1}  -(n+1)\,\mu_n=0\ ,
\ee
 whose solution is given in \eqref{munpara1} and \eqref{munpara2}. Next, one finds that 
 the consistency of the equation
 $\nabla^2\, + \left\{ \boldsymbol{\sigma}^- ,  \boldsymbol{\sigma}^+ \right\}
+ \left( \boldsymbol{\sigma}^0 \right){}^2   = 0$ gives the  recursion relation
\begin{eqnarray}
- n^{2}\, \lambda^2 &=& \frac{n^{2}}{(n+1)}\,
\frac{(n+\tfrac{3}{2})}{(n+\tfrac{1}{2})}\,\lambda_{n+1}  - n \, \lambda_n \ ,
\qquad \lambda_0=0\ ,~ \lambda_1=\tfrac{M^{2}_{0}}{3}\;,~  n \geqslant s=0\ ,
 \nonumber \\
 -\lambda^2 &=& \frac{\lambda_{n+1}}{(n+1)}\,
\frac{(n+\tfrac{3}{2})}{(n+\tfrac{1}{2})}  - \frac{\lambda_n}{n}
- \frac{\mu_n^2}{n^2}\ ,\qquad \lambda_s=0\;,\quad n\geqslant s>0\ ,
\end{eqnarray}
whose solution is given in \eqref{lambdan1} and \eqref{lambdan2}.
Finally, the equation $\left\{ \boldsymbol{\sigma}^+ ,  \boldsymbol{\sigma}^0 \right\} = 0$
is found to be satisfied as a consequence of the others, so that the integrability
conditions ${\cal D}^{2} = 0 $ have all been treated.

\end{appendix}


\providecommand{\href}[2]{#2}\begingroup\raggedright\endgroup


\end{document}